\documentclass[12pt]{amsart}
\usepackage{amsfonts,amssymb,amsmath,amscd}

\def\om{{\omega}}

\newcommand{\ZZ}{{\mathbb Z}}

\newcommand{\CC}{{\mathbb C}}

\def\cL{{\mathcal L}}
\def\be{\beta}

\def\Pf{{\rm Pf}}
\def\al{{\alpha}}
\def\sm{{\sigma}}
\def\bi{{\bar{i}}}
\def\bj{{\bar j}}
\def\bz{{\bar z}}

\def\Sm{{\Sigma}}
\def\Om{{\Omega}}
\def\Smc{{\Sigma_\CC}}
\def\tE{{\tilde E}}
\def\bound{{\part\Sm}}
\def\Ec{{E_\CC}}
\def\Ga{{\Gamma}}
\def\part{{\partial}}
\def\TX{{T\!X}}
\def\vol{{\rm vol}}

\def\End{{\rm End}}
\def\det{{\rm det}}
\def\bea{\begin{eqnarray}}
\def\eea{\end{eqnarray}}

\title[Anomalies and Graded Coisotropic Branes]{Anomalies and Graded Coisotropic Branes}
\author[Yi Li]{Yi Li}
\address{California Institute of Technology\\
Department of Physics\\
Pasadena, CA 91125}
\email{yili@theory.caltech.edu}

\date{May 2004}

\begin{document}

\begin{abstract}
We compute the anomaly of the axial $U(1)$ current in the A-model on a Calabi-Yau manifold, in the presence of coisotropic branes discovered by Kapustin and Orlov. Our results relate the anomaly-free condition to a recently proposed definition of graded coisotropic branes in Calabi-Yau manifolds. More specifically, we find that a coisotropic brane is anomaly-free  if and only if it is gradable. We also comment on a different grading for coisotropic submanifolds introduced recently by Oh.
\end{abstract}

\maketitle 

\vspace{-4in}

\parbox{\linewidth}
%{\small\hfill \shortstack{CALT-xx-xxxx}} 

\vspace{4in}

\section{Introduction}

Topological D-branes are important objects to study from both the physical and the mathematical point of view. Physically, they provide a simplified model for analyzing the boundary conditions in the full-fledged string theory. Frequently, they are simple enough to admit exact analysis and yet have rich enough structure to exhibit many common characteristics like their more mysterious siblings in superstring theory. A notable recent example  where topological D-branes play a prominent role is a large-$N$ duality proposed by Vafa \cite{Vafa}.

Mathematically, a major motivation to studying topological D-branes comes from the need to understand mirror symmetry. An $N=2$ sigma model on a Calabi-Yau manifold $X$ admits two inequivalent topological twistings. The resulting topological field theories are called the A-model and the B-model \cite{WittenTop}, and the D-branes in them are called topological A-branes and B-branes accordingly. On physical grounds, mirror symmetry exchanges the A-model on $X$ with the B-model on its mirror $\hat X$, and therefore must exchange the sets of A-branes and B-branes. One promising proposal to understand this mirror phenomenon in mathematical terms is the Homological Mirror Symmetry (HMS) conjecture \cite{Konts}, which interprets mirror symmetry as the equivalence of two triangulated categories: the bounded derived category of coherent sheaves $D^b(X)$ on the one hand, and the derived Fukaya category $DF(\hat X)$ on the other hand. It was later argued by Douglas \cite{Douglas} (see also \cite{AL}) that the derived category $D^b(X)$ corresponds to the category of topological B-branes. It is therefore tempting to regard the HMS conjecture as a mathematical re-phrasing of the physical statement that mirror symmetry exchanges A-branes and B-branes. One would then naively expect that the category of A-branes is the same as the derived Fukaya category, whose objects are $\ZZ$-graded Lagrangian submanifolds carrying flat vector bundles. In fact, it is known for a long time that Lagrangian submanifolds provides a prototype for topological A-branes \cite{WittenCS}. The notion of graded Lagrangian submanifolds, originally due to Kontsevich \cite{Konts} and later elaborated and generalized by Seidel \cite{Seidel}, is a refinement of ordinary Lagrangian submanifolds that turns out to be particularly significant physically as well.

There is an important modification to this story. It was noticed awhile ago that there might be non-Lagrangian submanifolds that may serve as A-type branes \cite{OOY}. However, the non-Lagrangian case had not received much attention until they were re-investigated carefully by Kapustin and Orlov \cite{KO}. It was found by these authors that, at the classical level, an important class of non-Lagrangian A-type boundary conditions are provided by certain coisotropic submanifolds carrying non-trivial line bundles, which we refer to as coisotropic branes. This finding suggests that the category of A-branes should be a suitable enlargement of the derived Fukaya category, with the coisotropic branes mentioned above providing primary candidates for the additional objects.

For a coisotropic brane to be a true topological A-brane, and thus an acceptable object in whatever extension of the Fukaya category, an additional anomaly-free condition must be satisfied. In the Lagrangian case, such an anomaly-free condition, analyzed first by Hori, turns out to be precisely that the Lagrangian submanifold be gradable in the sense of Kontsevich \cite{Hori}. This is a satisfactory result since it confirms the long-standing belief that the objects in the derived Fukaya category can indeed be regarded as topological A-branes. The anomaly-free condition for the coisotropic branes was not known previously, but it is natural to expect it to be associated with certain gradability condition by analogy with the Lagrangian case. In a recent paper \cite{KapLi}, a proposal for a possible definition of graded coisotropic branes is put forward based on a study of stability of A-type supersymmetric D-branes, and it is conjectured there that the gradability condition is the same as the anomaly-free condition for a coisotropic brane. It is the main objective of this paper to directly derive the anomaly-free condition for coisotropic branes, and our results prove this conjecture affirmatively.

This paper is organized as follows. In the next section, we summarize the essential geometric properties of coisotropic branes, and review the definition of the generalized Maslov class and the corresponding notion of graded coisotropic branes.  In Section \ref{sec:anomaly}, we derive the anomaly-free condition for coisotropic branes and relate it to the generalized Maslov class. In section \ref{sec:discussion}, we briefly comment on a different grading for coisotropic submanifolds introduced recently by Oh \cite{Oh}.

\section{Coisotropic Branes and Generalized Maslov Class}
\label{sec:cois}

In this section we review some basic facts about the geometry of coisotropic branes and their associated generalized Maslov class, based on the discussion in \cite{KO, KapLi}. An $N=2$ supersymmetric sigma model with boundary is defined by a map $\phi:\Sm\to X$ from the worldsheet $\Sm$ to a target space $X$, which we assume to be a Calabi-Yau manifold, i.e. a compact K\"ahler manifold with trivial canonical bundle. Denote the K\"ahler metric by $G$ and the K\"ahler form by $\om$. The bosons of the theory are given by the map $\phi$. The fermions of the theory are the left movers $\Psi_+\in\Ga(\phi^*\TX\otimes S_+)$ and right movers $\Psi_-\in\Ga(\phi^*\TX\otimes S_-)$, with $S_\pm$ being the spinor bundles on $\Sm$. For our purpose, a D-brane is a triple $(Y,\cL,\nabla)$, where $Y$ a submanifold of $X$ such that $\phi(\bound)\subset Y$, $\cL$ a line bundle on $Y$, and $\nabla$ a unitary connection on $\cL$. Let $F$ be the curvature of $\cL$, which is a real 2-form on $Y$. We will also use the notation $(Y,F)$ to refer to the D-brane defined by $(Y,\cL,\nabla)$. The boundary condition specified by $(Y,F)$ takes the form of $\Psi_+ = R\Psi_-$, where $R$ is a bundle map that can be represented in the following matrix form in a local basis with respect to the orthogonal decomposition $\TX|_Y\simeq TY\oplus NY$:
\begin{equation*}
R \;=\; \begin{pmatrix} -{\rm id}_{NY} & 0 \\ 0 & (G-F)^{-1}(G+F)|_{TY} \end{pmatrix}.
\end{equation*}

By definition,  a D-brane of type-A  is a boundary condition which preserves the sum of the left-moving $N=2$ super-Virasoro and the mirror of the right-moving $N=2$ super-Virasoro. In particular, this implies $R^tGR=G$ and $R^t\om R=-\om$. In the case of $F=0$, it is first shown by Witten that $Y$ must be a Lagrangian submanifold \cite{WittenCS}. The case of non-flat bundle is determined in \cite{KO}, whose results we summarize here. The first requirement is that $Y$ must be a coisotropic submanifold of $X$. This means that ${\rm ker}\,\om|_Y\equiv TY^\om\subset TY $ is an integrable distribution of constant rank in $TY$. Let $FY\equiv TY/TY^\om$, and note that the complex structure on $X$ naturally induces the decomposition $FY\simeq FY^{1,0}\oplus FY^{0,1}$. The second requirement says that the curvature 2-form $F$ of the line bundle annihilates $TY^\om$ and therefore descends to a section of $\wedge^2 FY^*$. Finally, $\om^{-1}F|_{FY}$ defines a transverse complex structure on $FY$. A direct consequence of the last condition is that $F_{0,2}$, the $(0,2)$-part of $F$, is non-degenerate. It follows easily from these conditions that the complex dimension of $FY$ must be even.

The analysis of \cite{KO} is carried out at the classical level. Quantum mechanically, a coisotropic brane is a topological A-brane if and only if an additional anomaly-free condition is satisfied. To explain this fact, recall that the A-model without boundary comes naturally with a $\ZZ$-grading, the charge of the axial $U(1)$ current. In particular, the topological correlators on the sphere preserve this $\ZZ$-grading, and this fact makes the bulk operator product algebra into a differential graded algebra. A topological A-brane must preserve this structure. In other words, it is necessary that the presence of the boundary does not break the axial R-symmetry. The coisotropic boundary condition found in \cite{KO} preserves the axial R-symmetry at the classical level, although it might induce a quantum anomaly  that spoils the $\ZZ$-grading of the theory. 

As already mentioned in Section 1, a Lagrangian brane is anomaly-free, and hence a topological A-brane, if and only if its associated Lagrangian submanifold is gradable in the sense of Kontsevich. For the coisotropic case, it is conjectured that a coisotropic brane is anomaly-free (and hence is a topological A-brane) if only it is gradable in the sense of a grading introduced in \cite{KapLi}. Let's briefly recall the relevant definition proposed in \cite{KapLi} here. Let $\Om$ be a holomorphic top form on the Calabi-Yau $X$ which is nowhere zero, and let $k\in 2\ZZ$ be the complex dimension of $FY$. As the $(0,2)$-part of  $F$ is non-degenerate, $\Om\wedge F^{k/2}$ is a nowhere vanishing top form on $Y$. Therefore one can write $\Om\wedge F^{k/2}|_Y = c\cdot\vol(Y)$, where $c:Y\to\CC^\times$ is a function to the punctured complex plane. Its logarithm $\log c$ is well-defined locally if one picks a (location-dependent) branch. However, there is an obstruction to lifting $\log c$ to a single-valued function globally, which is measured by a class in the Cech cohomology $H^1(Y,\ZZ)$. We define this obstruction class to be the generalized Maslov class of the coisotropic brane $(Y,F)$ and denote it by $\mu(Y,F)$. A coisotropic brane is called gradable if its generalized Maslov class is trivial. A graded coisotropic brane is a gradable coisotropic brane together with a global lifting of $\log c$, with its $\ZZ$-grading being a choice of the branch of $\log c$. 

As we will demonstrate in the following, this gradability condition is precisely the condition that the coisotropic brane be anomaly-free.  

\section{Anomalies of Coisotropic Branes}
\label{sec:anomaly}

In this section we derive the anomaly-free condition for coisotropic branes. As before, let $X$ be a Calabi-Yau manifold of complex dimension $n$, and let $(Y,F)$ be a coisotropic brane of real codimension $r=n-k$.  From the discussion in Section \ref{sec:cois}, $k$ must be an even integer. Let $\phi:(\Sm,\part\Sm) \to (X,Y)$ be the map that defines the worldsheet theory. Let $E=\phi^*\TX^{1,0}$ and $\bar E = \phi^*\TX^{0,1}$ be the pullbacks of the holomorphic and anti-holomorphic tangent bundles of $X$, and let $K$ denote the canonical bundle on $\Sm$. After the topological twisting, the fermions in the A-model are sections of the following bundles on $\Sm$:
$$\psi_+\in\Ga(E), \quad \psi_-\in\Ga(\bar E), \quad \rho_+\in\Ga(\bar E\otimes K), \quad \rho_-\in\Ga(E\otimes \bar K).$$
The K\"ahler metric on $X$ induces a natural hermitian metric on the pullback bundle $\phi^*\TX$, which we continue to denote by $G$. It will be convenient to write everything in a holomorphic basis with respect to the decomposition $\TX\simeq\TX^{1,0}\oplus\TX^{0,1}$ for carrying out explicit computation later. For example, the metric $G$ and the boundary map $R$ can be represented in the following matrix form under such a basis:
\begin{equation}
\label{eq:R}
G = \begin{pmatrix} 0 & g \\ g^t & 0 \end{pmatrix}, \qquad R = \begin{pmatrix} 0 & R_a \\ R_b & 0 \end{pmatrix}.
\end{equation}
Explicit expression for $R$ can be found in \cite{KapLi}. 

The kinetic action of the fermions looks like
$$\sqrt{-1}\cdot \int_\Sm \;G(\rho_+, D_\bz\psi_+) + G(\rho_-, D_{z}\psi_-)$$
where $D_z$ and $D_\bz$ are covariant derivatives defined by the pullback of the Levi-Civita connection on $\TX$.  Under the axial R-symmetry, $\psi_\pm$ have charge $+1$ while $\rho_\pm$ have charge $-1$. This is a symmetry of the bulk A-model because of the Calabi-Yau condition. The coisotropic branes discussed in Section \ref{sec:cois} preserve the axial R-symmetry at the classical level. 

Any potential anomaly in the axial R-symmetry must come from the zero modes of the fermions. More specifically, we must compute the following index
$$ \mbox{\# $(\psi_+,\psi_-)$ zero modes} \;-\; \mbox{\# $(\rho_+,\rho_-)$ zero modes}$$
subject to the boundary conditions
$$\psi_+=R_a\psi_-, \qquad \rho_+=R_b\rho_-.$$
In order to enumerate the zero modes, we use a doubling trick that effectively converts the problem to an index theorem on a compact Riemann surface. Such doubling methods have been used in recent studies of Lagrangian boundary conditions in \cite{KL,Hori}. The basic idea is to double the worldsheet $\Sm$, and to interpret $(\psi_+,\psi_-)$ and $(\rho_+,\rho_-)$ as fields propagating on the doubled surface. Mathematically, this means that we want to interpret them as sections of certain complex vector bundle defined on the doubled surface. For simplicity, we assume $\bound\simeq S^1$ in the following discussion, although the result of our analysis does not depend on this fact in any essential way.

Let's choose a metric on $\Sm$ which is a cylindrical product around $\part\Sm$. Its orientation reversal, denoted by $\Sm^*$, carries the opposite complex structure and has a metric naturally induced by that on $\Sm$. Using the metric, one can glue $\Sm$ and $\Sm^*$ along $\part\Sm=\bound^*$, yielding a compact Riemann surface $\Sm_\CC$. We call $\Smc$ the complex double of $\Sm$. Let $\sm:\Sm^*\to\Sm$ be the reflection map, and let $\tE=\sm^*\bar E$ be the pullback bundle of $\bar E$. The crucial idea then is to regard $\psi_-$ and $\rho_-$ as fields living on $\Sm^*$, as in \cite{Hori}. The precise meaning of this is that one identifies $\psi_-$ with its pullback section in $\tE$. Similarly, one identifies $\rho_-$ with its pullback section in $\tE\otimes K^*$, where $K^*$ is the canonical bundle of $\Sm^*$. In the following, we shall construct a complex vector bundle $\Ec\to\Smc$ such that the pair $(\psi_+,\psi_-)$, when properly patched together by the boundary condition, define a smooth section on it.

Since $\bound$ is non-empty, the pullback bundle $\phi^*\TX$ is trivial. Fixing a trivialization of $\phi^*\TX$ induces canonical trivializations 
$$\varphi:E\to\Sm\times\CC^n, \qquad \varphi':\bar{E}\to\Sm\times\CC^n.$$ 
Note that $\varphi'$ naturally induces a trivialization of $\tE$, which we also denote by $\varphi'$ by a slight abuse of notation. We point out that sections of $E$ and $\tE$ are trivialized by $\varphi$ and $\varphi'$ with respect to {\em conjugate} bases of $\CC^n$. When one represents $\psi_+\in\Ga(E)$ and $\psi_-\in\Ga(\tE)$ in the component form 
$$\psi_+ = \psi_+^i e_i, \qquad \psi_- = \psi_-^\bi {e}_\bi$$
it is implicit that such a trivialization pair $(\varphi,\varphi')$ are chosen, with $\{e_i\}$ and $\{e_\bi\}$ being conjugate bases. So is the case when one writes the boundary condition $\psi_+=R_a\psi_-$ in the matrix form (see (\ref{eq:R})):
\begin{equation}
\label{eq:Ra}
\psi_+^i=(R_a)^i_{\;\bj}\,\psi_-^\bj\,.
\end{equation}
The reason that we elaborate on this seemingly trivial fact is that, for the purpose of constructing $\Ec$, it is essential that one trivializes the bundle over different patches with respect to the same basis of $\CC^n$. This suggests that a more natural trivialization of $\tE$, for our purpose, is actually the conjugate of $\varphi'$:
$$\bar\varphi':\;\tE\to\Sm^*\times\CC^n.$$

After these preliminary remarks, we are ready to construct $\Ec$ by, roughly speaking, gluing $\tE$ and $E$ along $\bound$. Take an open covering $\{U_\al,U_\be\}$ of $\Smc$ with $\Sm\subset U_\al$ and $\Sm^*\subset U_\be$, such that $U_{\al\be}=U_\al\cap U_\be$ is a tubular neighborhood of $\part\Sm$. Let's extend $E$ and $\tE$ to $U_\al$ and $U_\be$ respectively. It is tempting to let $\Ec$ be the vector bundle whose trivializations over $U_\al$ and $U_\be$ are simply given by $\varphi_\al=\varphi$ and $\varphi_\be=\bar\varphi'$. This is problematic since it does not take into consideration of the boundary condition, and one can easily check that $(\psi_-,\psi_+)$ do not define a smooth section of the bundle $\Ec$ constructed this way. 

It is not difficult to remedy the problem. Our construction amounts to interpreting $R_a$ as an endomorphism of $\tE$ and the boundary condition (\ref{eq:Ra}) as a transition function. To this end, let's extend $R_a\in\Ga(\End\tE|_\bound)$ to $U_{\al\be}$. Such an extension always exists. By using a bump function, one can actually extend $R_a$ to the whole of $U_\be$, such that it is nowhere degenerate, and it becomes the identity endomorphism outside of a small open neighborhood of $U_{\al\be}$. By yet another slight abuse of notation, we use the same symbol $R_a$ to denote its extension to $U_\be$. 

To complete the construction of $\Ec$, we take $\varphi_\al = \varphi$ as before and take $\varphi_\be=\bar\varphi'\circ R^{-1}$. This gives the desired bundle $\Ec$, whose transition function with respect to the open cover $\{U_\al,U_\be\}$ is given by
\begin{equation}
\label{eq:h}
h_{\al\be} = \varphi_\al\cdot\varphi_\be^{-1} = \varphi\cdot R\cdot\bar{\varphi'}^{-1}.
\end{equation}
In particular, one can check that $\psi_+$ and $R_a\psi_-$ glue smoothly into a single section $\chi\in\Ga(\Ec)$, with the gluing condition restricted to $\bound$ being precisely the boundary condition (\ref{eq:Ra}). Similarly one can show that $\rho_+$ and $R_b\rho_-$ glue smoothly into a section $\eta\in\Ga(\Ec\otimes K_\CC)$, with $K_\CC$ being the canonical bundle of $\Smc$.

It remains to relate the problem of counting zero modes to an index theorem on $\Smc$. As $E$, $\tE$ are trivial bundles, their connections are simply endomorphism-valued 1-forms. Using the invariance property of the index, we can pick any connections on $E$ and $\tE$, as long as the boundary condition is preserved in a covariant way. From this point of view, the fermionic action can equivalently be written as, in terms of the global fields $\chi$ and $\eta$:
$$\sqrt{-1} \int_\Smc \; G(\eta,\bar\part_A\chi).$$
Here $\bar\part_A$ is a twisted Dolbeault operator and $A$ is a connection on $\Ec$ which we might as well take as
$$A|_{U_\al} = 0, \qquad A|_{U_\be} = h_{\al\be}^{-1}d h_{\al\be}.$$
Therefore we have converted the problem of enumerating the difference of the numbers of $(\psi_+,\psi_-)$ zero modes and the $(\rho_+,\rho_-)$ zero modes into calculating the index of a twisted Dolbeault operator associated with the complex vector bundle $\Ec$, which by a well-known index theorem is given by
$${\rm ind}\, \bar\part_A \;=\; c_1(\Ec) + n(1-{g}_\CC)$$
with $g_\CC$ being the genus of the doubled surface $\Smc$. In the formula above, the second term on the RHS is a non-anomalous contribution, since it is a topological constant that does not depend on details of the map $\phi:(\Sm,\bound)\to (X,Y).$ If nonzero, it simply shifts the ``ghost number'' of the vacuum state. The first term, on the other hand,  depends on the map $\phi$ explicitly. If nonzero, there is no consistent way to assign a $\ZZ$-grading to the operators of the theory using the axial $R$ charges. Based on earlier discussion, we conclude that a coisotropic brane $(Y,F)$ is a topological A-brane if and only if $c_1(\Ec)=0$ for any map $\phi:(\Sm,\bound)\to(X,Y)$. 

Let's relate the anomaly $c_1(\Ec)$ to the generalized Maslov class defined earlier. The crucial link is provided by a holomorphic top form $\Om$ that is nowhere vanishing on $X$. Let $w:\bound\to Y$ be the restriction of $\phi$ to $\bound$. Over $w(\bound)\subset Y$, the tangent bundle $\TX$ is trivial for topological reasons. One can then choose a unitary frame on $w(\bound)$
$$u_i \;=\; \frac1{\sqrt2}\big(e_i+\sqrt{-1}f_i\big), \qquad i=1,2,\ldots,n$$
such that $\{u_1,u_2,\ldots,u_{k}\}$ and their conjugates span $FY$. In addition, we can assume $\{e_{k+1},\ldots,e_n\}$ span $TY^\om$. With respect to this frame, one has
$$\Om\wedge F^{k/2}|_{w(\bound)} \;=\; \Om_{12\ldots n}{\rm Pf}(F_{0,2})\cdot{\rm vol}(Y)$$
up to a normalization constant. Here $\Om_{12\ldots n}$ is the contraction of $\Om$ with $u_1\wedge\cdots\wedge{u}_n$, and ${\rm Pf}(F_{0,2})$ is the Pfaffian of the $(0,2)$-part of the 2-form $F$. This shows that $\Om_{12\ldots n}{\rm Pf}(F_{0,2})$ is just the function $c:Y\to\CC^\times$ appearing in the definition of the generalized Maslov class of $(Y,F)$, restricted to $w(\bound)$.

Using the pullback map $w^*$, one obtains a section $w^*(u_1\wedge\cdots\wedge u_n)$ of $\wedge^n E$ over $\bound$, which can be extended to a local section $s$ of the determinant bundle $\wedge^n\Ec$ over $U_{\al\be}$.  By contraction, $\phi^*\Om$ provides a trivialization for $\wedge^n\Ec|_{U_\al}$, under which the component of $s$ is simply $\phi^*\Om_{12\ldots n}$. Under the trivialization on the patch $U_\be$, the same section is mapped to $\phi^*\big(\Om^*_{12\ldots n}/\det(R_a)\big)$. This gives a concrete realization of the transition function for the determinant bundle $\wedge^n\Ec$:
$$\det (h_{\al\be}) \;=\;\phi^*\left( \frac{\Om_{12\ldots n}}{\Om^*_{12\ldots n}}\cdot\det(R_a)\right).$$
Restricting to $\bound$, it defines a function $S^1\to\CC^\times$ whose winding number is the first Chern number $c_1(\Ec)$. As is shown in \cite{KapLi}, the determinant of $R_a$ takes the following form
$$\det (R_a) \;=\; \det(F_{0,2})/\det(\tilde{g}-F_{1,1})$$
where $\tilde{g}$ is the restriction of the metric to $FY$, and $F_{1,1}$ is the matrix associated with the $(1,1)$-part of $F$. Since the denominator is real, it does not contribute to the winding number of $\det(h_{\al\be})$. Therefore $c_1(E_\CC)$ is twice the winding number of $\phi^*(\Om_{12\ldots n}\cdot{\Pf}(F_{0,2}))$. From the discussion in the last paragraph, we conclude that the first Chern number $c_1(\Ec)$, which measures the anomaly in the axial $U(1)$ current, is given by
$$c_1(\Ec) \;=\; 2\langle \phi^*\mu(Y,F), \al\rangle$$
where $\al$ is the generator of $H_1(\bound\simeq S^1,\ZZ)$. 

As anticipated, the anomaly-free condition for the axial R-symmetry in the presence of a coisotropic brane is that the generalized Maslov class of the brane be trivial, i.e.~the coisotropic brane is gradable in the sense of \cite{KapLi}.

\section{Discussion}
\label{sec:discussion}

In this concluding section, we would like to comment on a potentially confusing issue concerning different definitions of grading associated with coisotropic submanifolds. Recently, a definition for graded coisotropic submanifolds is proposed by Oh \cite{Oh}. We briefly recall Oh's definition below. Let $X$ be a symplectic manifold with a compatible (almost) complex structure $J$, and let $Y\subset X$ be a coisotropic submanifold. The almost complex structure $J$ naturally decomposes $FY\equiv TY/TY^\om$ into $FY^{1,0}\oplus FY^{0,1}$. The transverse canonical bundle $K_Y$ of the coisotropic submanifold $Y$ is defined to be the determinant bundle of $(FY^*)^{1,0}$. According to Oh's definition, $Y$ is a gradable coisotropic submanifold if $K_Y^{\otimes2}$ is trivial, and a graded coisotropic submanifold is a gradable coisotropic submanifold with a global section of $K_Y^{\otimes2}$. 

It is not difficult to see the essential differences between Oh's definition of gradable coisotropic submanifolds and our definition of gradable coisotropic branes. Most importantly, Oh's definition is intrinsic to the almost K\"ahler structure $(X,\om,J)$, while our definition involves additional structure associated with the gauge field living on $Y$. Every coisotropic brane $(Y,F)$, gradable or not according to our definition, is a graded coisotropic submanifold in the sense of Oh, with $F_{2,0}^{k/2}$ providing a global section of $K_Y$. While Oh's definition applies to more general situations\footnote{Recall that the definition in \cite{KapLi} works only when $X$ is Calabi-Yau.} and is certainly an interesting geometric construction, it is not what one needs for characterizing topological A-branes. Indeed, as already mentioned in \cite{Oh}, Oh's definition of graded coisotropic submanifolds is not a generalization of the notion of graded Lagrangian submanifolds defined by Kontsevich and Seidel. In fact, it is obvious that every Lagrangian submanifold is gradable in the sense of Oh. The result of this paper suggests that it is the graded coisotropic brane defined in \cite{KapLi} that provides a proper generalization of the graded Lagrangian submanifold from the point of view of both topological field theory and categorical mirror symmetry.

\section*{Acknowledgments}
I am grateful to Anton Kapustin for illuminating discussions and for carefully reading the manuscript and making numerous suggestions.  I am also indebt to Ke Zhu for helpful discussions and for reading a preliminary version of the manuscript. This research is supported in part by DOE grant DE-FG03-92-ER40701.

\end{document}